\documentstyle[12pt]{article}
\begin{document}
\newcommand{\tr}{{\rm {\bf {t\!r}}}}
\typeout{--- Title page start ---}

\thispagestyle{empty}
\renewcommand{\thefootnote}{\fnsymbol{footnote}}

\begin{tabbing}
\hskip 11.5 cm \= {Imperial/TP/93-94/14}\\
\> hep-th/9402062 \\
\> December, 1993 \\
\> To appear in the JMO\\
\end{tabbing}
\vskip 1cm
\begin{center}
{\Large\bf Some Remarks on Quantum Coherence\footnote{
To appear in the {\em Journal of Modern Optics}
special issue on quantum communication.
}
} \\
\vskip 1.2cm
{\large\bf Andreas Albrecht}\\
Blackett Laboratory, Imperial College\\
Prince Consort Road, London SW7 2BZ  U.K.\\
albrecht@ic.ac.uk
\end{center}
\vskip 1cm
\begin{center}
{\large\bf Abstract}
\end{center}

There are many striking phenomena which are attributed to
   ``quantum coherence''.   It is natural to wonder if
there are  new quantum coherence effects waiting
to be discovered which could lead to interesting results and
perhaps even practical applications. A useful
starting point for such discussions
is a definition of ``quantum coherence''.   In this article
I give a definition of quantum coherence and use
a number of illustrations to explore the implications of this definition.
I point to
topics of current interest in the fields of cosmology and
quantum computation where questions of quantum coherence arise, and I
emphasize the  impact  that interactions with the
environment can have on quantum coherence.

\vskip 1cm
\typeout{--- Main Text Start ---}

\renewcommand{\thefootnote}{\arabic{footnote}}
\setcounter{footnote}{0}

\section{Introduction}

In a double slit experiment, the striking pattern which appears on the
photographic plate is different from anything one could expect from
classical ``particle''.   This is but one of many striking phenomena
associated with
``quantum coherence''.   It is very interesting to consider the
possibility that some startling new quantum coherence effects are
waiting to be discovered.  I am familiar with  two diverse fields where
this issue has come up, in early universe cosmology and quantum
computation.

In cosmology, Grishchuk and collaborators have pointed out that a
period of cosmic inflation will naturally land the universe in a coherent
superposition of classical
states\cite{grishchuk91q,grishchuk92,grishchuk&sidorov90}.
It would be very interesting if there were astrophysically observable
effects of this quantum coherence.
 My collaborators and I have argued, contrary to claims in
the literature,  that no calculations so
far have revealed quantum coherence effects, and that it is unlikely
that such effects could be
observed\cite{albrecht&ferreira&joyce&prokopec93}.
   Still, it would be extremely
interesting if our pessimism proved to be wrong.

In the field of quantum computation,  one asks whether anything
interesting, or
perhaps even useful, can be gained by allowing computers to evolve
as  coherent superpositons of computational
states\cite{feynman85,deutsch85,deutsch89}.
  The issues in
computation theory are fascinating and recent progress has
generated a great deal of excitement.  As a practical matter, any
attempt to engineer such a computer would run up against many of
the same issues we are confronting in cosmology.

A useful starting point for any of these discussions
is a definition of quantum coherence, and that is the first topic
of this article.  The word ``coherence'' has been  used in a number of
different contexts. It is worth noting that here I am not specifically
concerned with the notion of the ``coherence of a field'' as
developed, for example, by Glauber \cite{glauber65}.  Although not
completely unrelated, the notion of ``quantum coherence'' used here
will be a
different one (which I define precisely in Section 2).

Regular readers of JMO may find sections 1-4 elementary.  These
experts might,
however, be interested in the way these  familiar issues arise in
cosmology and quantum computation.   This article is also directed at
the diverse community of scientists who have developed an interest in
quantum
computers.  The introductory material should be useful
to some of these readers.

This is not an article about the foundations of quantum mechanics.
The points discussed here may be equally well viewed from a
Copenhagen or Everett viewpoint (and no doubt from other viewpoints as
well).
The points made in this article could probably be
re-expressed quite elegantly in the
``consistent histories'' formalism, but aside from briefly mentioning
this possibility
in Section \ref{ch} I do not pursue this avenue here.

Section 2 defines quantum coherence, and explores the implications
of this definition.  The double slit experiment and the WKB states are
taken
as illustrations, and I also make some general remarks about the {\em
subjectivity} of quantum coherence.   The WKB case is peculiar in that
on one hand it is the most classical type of state. On the other hand
 the crucial role played by the complex phase in relating the quantum
Hilbert
space to the classical phase space allows one to take the unusual
view
that all of classical physics is a ``quantum coherence effect''.

Section 3 examines the important role played by interactions with the
``environment''.  Such interactions are ignored in section 2, and the
discussion in Section 3 shows how dramatically the picture can change
when
they are included.  Section 4 relates the previous discussion to
current issues
in ``quantum computation'', and section 5 describes the way in which
questions
of quantum coherence have recently turned up in my own field of
early universe cosmology.  Section 6 briefly points out possible links
between
the content of this paper and
the ``consistent histories'' formalism. Section 7 contains the
conclusions.

\section{Quantum coherence in isolated systems}

\label{qciis}

\subsection{Definition of quantum coherence}

\label{doqc}

Classical mechanics invites one to view a physical system as point in
the ``phase space'' of possible states.  In practice one is
ignorant about many
details of the physical state of the system. The space of microscopic
states is typically enormous, and instead of
considering a single point in phase space, one assigns probabilities to
a range of possible physical states.  Based on the dynamics of the
system, one can evolve an initial probability
distribution into a ``final'' one,
representing what we know about the system at some later time.

Quantum mechanics is different from classical mechanics in several
ways.  Firstly, the state of a system is defined most fundamentally by
probability {\em amplitudes} (the ``wavefunction'') which must be
squared to get the probabilities.  Secondly, the space of possible
quantum states
is quite different from its classical
counterpart. Positions and momenta can not be both specified
precisely, the spectra of bound states  tend to be discrete, etc.  .

This article is concerned with the extent to which quantum systems
{\em can}  be treated using  probabilities instead of probability
amplitudes.   To the extent that the probabilities
are all one needs,  I will say one is
working with a ``classical'' probability distribution,
regardless of whether the actual space of possible
states has quantum mechanical features or not.
\begin{verse}
{  To the extent that one needs to know the
initial probability {\em amplitudes} (rather than
just the probabilities)
in order to do the right
calculation, I will say that the system exhibits ``quantum coherence''}.
\end{verse}

The difference between systems which can be described with
classical probabilities and those which exhibit  quantum coherence is
a practical one.   Any physicist should be aware of this
distinction, since the type of calculation one undertakes
can be very different in the two cases.
It is possible that not everyone would agree
that this distinction corresponds to {\em their}
notion of quantum coherence, but I am not interested
in arguing this point.  The distinction I choose to make
here is an important one in any case.  I suspect most readers
who study the illustrations below will
find this definition coincides with their notion of quantum coherence.

\subsection{Choice of basis}

\label{cob}

A state $|\psi\rangle $ for a quantum system can be expanded in any
orthonormal basis.  One gets something of the form
\begin{equation}
|\psi\rangle = \sum_i \alpha_i |i\rangle.
\label{expand}
\end{equation}
The complex numbers $\alpha_i$ are the probability {\em
amplitudes} for the system to be found in state $|i\rangle $, and the
positive real numbers
\begin{equation}
p_i \equiv \alpha^{\star}_i \alpha_i
\label{pdef}
\end{equation}
are the probabilities.  If one never used anything but the $p_i$'s, the
discussion would be indistinguishable
{ }from a discussion of a {\em classical }
probability distribution.

The interesting thing about quantum mechanics is that one can
discuss the state $|\psi\rangle$ in {\em any} basis.  Denoting a
different orthonormal basis with primes:
\begin{equation}
|\psi\rangle = \sum_i \alpha'_i |i\rangle'
\label{expandprime}
\end{equation}
with
\begin{equation}
\alpha'_i = \sum_j \alpha_j\;{ }'\!\langle i| j\rangle
\label{aprimedef}
\end{equation}
giving
\begin{equation}
p'_i = \sum_{j,k} \alpha^{\star}_j \alpha_k\; { }'\!\langle k | i \rangle
\langle i | j \rangle'.
\label{pprime}
\end{equation}
Clearly one needs to know the $\alpha_i$'s, not just the $p_i$'s, in
order to determine the $p'_i$'s.
Using the definition from Section \ref{doqc}, it would appear that
calculations which
requires a change of
basis involves quantum coherence.

\subsection{Time evolution}
\label{te}

When the evolution of $|\psi\rangle$ over a particular time interval
is given by the unitary operator $T$, Eqn \ref{expand}
becomes
\begin{equation}
T|\psi\rangle = \sum_i \alpha_i T|i\rangle.
\label{expandt}
\end{equation}
Now  $p_i = |\alpha_i|^2 $ gives the probability for the system to be
found in the state $T |i \rangle$.  If this is what one wants to
know, one does not need to use the complex phases of the $\alpha_i$'s,
and one is working with classical probabilities.  If  one is going to
measure the system in some other basis, one will need the complex
phases of the $\alpha_i$'s to determine the probabilities assigned to
the new basis states.  The $p_i$'s alone will not provide enough
information and one will encounter quantum coherence.

\subsection{An Illustration: WKB states}

\label{aiws}

Consider the wavefunction $\psi(x)$ for a single particle.

The
 wavefunction can be written
\begin{eqnarray}
\psi(x)=\rho(x)e^{iS(x)}.
\label{wkbform}
\end{eqnarray}
If $S(x)$ varies much more rapidly with $x$ than
$\rho(x)$ the state is a WKB state for which
\begin{eqnarray}
\hat{p}|\psi \rangle \simeq (\hbar \partial_x S(x))|\psi \rangle.
\label{ppsi}
\end{eqnarray}
To the extent that
this ``WKB condition'' holds
the state assigns momentum and position simultaneously
according to
\begin{eqnarray}
p(x)=\hbar \partial_xS(x).
\label{poq}
\end{eqnarray}
Using this expression one can construct from $\psi(x)$ a
probability distribution $P(x,p)$ in {\em classical} phase
space, in which both $p$ and $x$ are specified:
\begin{equation}
P(x,p) \equiv \psi ^{\star}(x)\psi (x)\delta (p - p(x)).
\label{wkbpxdist}
\end{equation}
It can be shown that to the extent that the WKB
condition continues to hold, this probability
distribution evolves according to
the equations for a distribution of classical particles
\cite{diracbook}.  Thus the time evolution of $P(x,p)$ can be
determined simply by solving these classical equations, given the initial
$P(x,p)$.    The initial probability
distribution is all one needs to determine the probability
distribution at later times.

Does a WKB system exhibit quantum coherence?  On one hand the
evolution can be
described entirely in terms of classical probabilities (not probability
{\em amplitudes}).  On the other hand,
Eqs \ref{poq}  and \ref{wkbpxdist} clearly show that any prediction
depends
quite explicitly on the complex phase $S$ of
the wavefunction ({\em not} just the probability $\rho^2(x)$).  The
subtlety  lies in the fact that the probability
distribution $P(x,p)$ which behaves nice and classically is not simply
a set of quantum probabilities $\rho^2(x)$.
 The evolution of
$P(x,p)$ depends very much on the complex phase $S$ of the wavefunction, but
this dependence can be re-expressed (using Eq \ref{poq}) in terms of a
classical
quantity (the momentum).  Thus one is dealing with classical probabilities
{\em and} complex phases at the same time.

Using a strict interpretation of the definition of quantum coherence given
in section \ref{doqc}, one would have to say that a  WKB system exhibits
a high degree of quantum coherence.  One is clearly making use of
the complex phase $S$, not just the probability $\rho^2(x)$\footnote{
When one studies the Wigner function the quantum coherence of WKB
states manifests itself in an interesting way:  The Wigner function
for WKB states does {\em not} simply give the appropriate classical
probability distribution.  The Wigner function is in fact negative in
some places, signaling the importance of quantum coherence effects
\cite{berry77,anderson90,habib90}
}.
  None the less, one can think of the system purely in terms of the
classical
probabilities $P(x,p)$.  For this reason it might be more in the {\em
spirit}  of the definition to say that a WKB system does not exhibit
quantum coherence.

Despite the ways in which WKB states are classical
there is another
 sense in which WKB states can be very quantum mechanical.
Quantum mechanics can best represent a classical particle with
a wave packet which is reasonably well localized in both $p$ and $x$.
For WKB states $P(x,p)$ need not be at all localized  in $x$ and $p$.
A delocalized  WKB state is a coherent superposition of many
classical wave packets.  At the end of the section \ref{aitdse}
I will discuss an example of highly delocalized WKB
states: ``squeezed states''.

\subsection{An Illustration: The double slit experiment}

\label{aitdse}

Consider the state $|\psi(t_1)\rangle$ (or
wavefunction $\psi(x,t_1) \equiv \langle x| \psi(t_1)\rangle$)
of an electron just as it is
passing through a barrier with double slits. The quantity $P(x,t_1)
\equiv
\psi^{\star}(x,t_1)\psi(x,t_1)$ gives the probability density for finding
the
electron at point $x$.
The wavefunction $\psi(x)$ is the continuum
equivalent to the $\alpha_i's$, with position eigenstates
$|x\rangle$ ( or delta
functions) as the basis.   Consider the unitary operator
$T$ which evolves the wavefunction to  a time $t_2$ when the
electron is well past the barrier.  The probability to find the
electron at position $x$
\begin{equation}
P(x,t_2) \equiv \left| \langle x\right|T\left|\psi(t_1)\rangle \right| ^2
\label{px}
\end{equation}
will reflect the special double slit
diffraction pattern, which is commonly
viewed as a quantum coherence effect.  (For electrons with momenta
comparable to $\hbar / a$, where $a$ is the slit width, this is
certainly {\em not} a WKB system.)

It is interesting to note that {\em if} one were to choose to
measure the electron at time $t_2$ in the basis given by $T|x\rangle$
(for each $x$) the probabilities would be:
\begin{equation}
P(Tx,t_2)  = \left| \langle x\right|T^{\dagger}T\left|\psi(t_1)\rangle
\right| ^2 = P(x,t_1).
\label{pTx}
\end{equation}
The complex phases of the initial wavefunction would be irrelevant,
and there would be no quantum coherence effects.
Furthermore, having determined which state $T|x\rangle$ the electron is
in at time $t_2$, one would know exactly where the electron had been at $t_1$
(but {\em only} at $t_1$!).
Our famous inability to determine which slit the electron passed through is
an ``artifact'' of our insistance on measuring the electron {\em position}
at late times, rather then measuring in the $T|x\rangle$
basis\footnote{
Knowing the location at $t_1$ does not actually tell you which slit
the electron has {\em passed through} since it provides no
information about the momentum of the electron.
To determine this one would best use a ``wave packet''
type basis at $t_1$, which would of course be just as
difficult to measure at $t_2$ as the $T|x\rangle$ basis.
}.
As a
practical matter the $T|x\rangle$ states are highly spread out and
differ from each other in highly non-local ways.  The prospect of
building a device for measuring these states is certainly remote
if not  completely out of the question.

In contrast, for  WKB states our tendency to measure classical
positions or momenta allows us to ignore questions of quantum
coherence
Probabilities can be assigned to the outcome of such
measurements using only the
{\em   probabilities} at an earlier time.
(Strictly speaking the complex phase is used, but interpreted
in a very
classical way.)
However, one need not always make classical measurements of WKB states.
In the laboratory, electromagnetic radiation can be put into
``squeezed states'' which are classical in the WKB sense.  To the extent
that photodetectors measure something other than the classical field
variables (number eigenstates of the field modes for example) one can
in principle observe some dramatic quantum coherence
effects\cite{schleich&wheeler87}. (That is,
effects for which the impact of the complex phase $S$ can {\em not} be
simply
expressed in terms of a classical momentum.)  Ref
\cite{zhu&caves90}  discusses some practical limitations on such
experiments which limit the observability
of these effects.
SQUIDs are another  example of places where quantum coherence effects
may be  observed
for large WKB states (see for example \cite{scpw92}).

If the  theory of cosmic inflation is correct, Grishchuk has pointed
out that all the matter in the universe was once in a squeezed state.
It appears to be  unlikely that in this case that the effects of
quantum coherence  (other than those corresponding to the classical WKB
probability distribution)
can be observed (see Section \ref{ssic}).

\subsection{The subjectivity of quantum coherence}

\label{tsoqc}

We have seen how the presence of quantum
coherence is related to what way one chooses to measure a system.  In
principle a suitable basis may always be chosen so that no quantum
coherence is observed, but in practice this choice may turn out not
to be realistic.

I now point out that quantum coherence, as defined in this paper,
also depends on how one specifies the initial state.  After all,
it is always possible to choose a basis for which the first basis
state $|1\rangle$ is  exactly equal to the initial state of the
system in question.  Then using only information about the initial
probabilities (namely $p_1 = 1$) one can determine {\em anything}
about the future of the system, no matter in  what basis the next
measurement will be made.

Thus the discussion of quantum coherence really has to do with
what bases we are inclined (or able) to use in describing a system at two
different times.   The extent to which quantum coherence effects
are observed depends on the relationship between these two bases.
To the extent that there is sufficient freedom to choose the bases, it will
be feasible to describe the system in a way that does not involve
quantum coherence.

\section{Quantum coherence in non-isolated systems}

\label{qcinis}

\subsection{Correlated states}

\label{cs}

The  interaction of one quantum system with another can
greatly affect questions of quantum
coherence
\cite{z82,c&l83,j&z85,walls&milburn85,z86,u&z89,z91,a92q,paz&habib&zurek93r}.
These interactions
will generically occur, since one is almost always
concerned with the behavior of a {\em subsystem} of
the whole universe, and the subsystem must be not completely isolated
{ }from the rest for it to be of any interest at all!   Thus, instead of
Eq \ref{expand}, one should write:
 \begin{equation}
 \left|\psi \right\rangle_{w} = \sum_{i,j} \alpha_{i,j}
 \left|i\right\rangle_{s}\otimes
 \left|j\right\rangle_{r},
 \label{genu}
 \end{equation}
Where  $ \left|\psi \right\rangle_{w}$ is the state of the ``world'',
and the orthonormal bases $\{\left|i\right\rangle_{s} \}$ and
$\{ \left|j\right\rangle_{r}\} $ span the ``system'' and the ``rest''
respectively.

If it were possible to write the initial state as
 \begin{equation}
\left|\psi \right\rangle_{w} = \left|\psi\right\rangle_s\otimes
\left|\psi \right\rangle_r,
 \label{sepu}
 \end{equation}
and if the the system were sufficiently isolated (until it was
measured), then it would be reasonable to ignore the ``rest'' and
study only the evolution of the system state
$\left|\psi\right\rangle_s$.  This would correspond to the discussion
in Section \ref{qciis}.

In the more general case ($\left|\psi \right\rangle_{w}$ given by Eq
\ref{genu}) it is not even possible to
assign a pure state to the system subspace.  Instead the system state
is described by the density matrix $\rho_s \equiv \tr _r |\psi\rangle_w {
}_w\!\langle \psi |$.  It is possible, however,
to choose the ``Schmidt'' basis\cite{s07}, which brings a general
state given by
Eq \ref{genu} into the
following more illuminating form:
 \begin{equation}
 \left|\psi  \right\rangle_{w} = \sum_i \sqrt{p_i^S}
 \left|i\right\rangle_{s}^S\otimes
 \left|i\right\rangle_{r}^S.
 \label{sumsepu}
 \end{equation}
Note that there is only {\em one} index being summed, and each basis
state appears in but one term.
The special Schmidt bases for the $r$ and $s$ subspaces are just
the eigenstates of the corresponding density matrices $\rho _r$ and
$\rho _s$.  The
$p_i^S$'s
are the eigenvalues of the density matrices.  (Both $\rho_r$ and
$\rho_s$ have identical eigenvalues,
with the larger one having additional zero eigenvalues.)  For
completeness,  I give a
simple introduction to  the Schmidt result in Appendix A. This
result has seen a large number of fruitful applications in
a variety of areas
where correlations are important
\cite{s35,z73,p&k91,gisin91,btt&phx91,a92q,btt&phx92,peres93,h&e93,aja94o}.

Given Eq \ref{sumsepu} the probability assigned to any system state
$|x\rangle_s$ is
\begin{eqnarray}
p_x & = & \sum_{ij} \sqrt{p_i p_j} { }_s\!\langle x | i\rangle_s^S
{ }^S\!\langle j | x\rangle _s { }_r^S\langle j | i\rangle_r^S
\\
& =  & \sum_i p_i^S \left|{ }_s\langle x| i\rangle_s^S \right| ^2.
\label{expx}
\end{eqnarray}
Thus the Schmidt basis is like the special basis discussed in Section
\ref{tsoqc}:  One need know nothing more than the {\em probabilities}
assigned to the Schmidt states to calculate anything about the system.
The Schmidt basis actually coincides with the special basis of Section
\ref{tsoqc} in the limit when Eq \ref{sepu} holds.

\subsection{Time evolution of correlations}

\label{teoc}

In the general case of a system interacting with its environment, one
needs to know not just the state of the system, but the state of the
environment as well.  In general there is nothing predictable about
the time evolution of the system state alone.  This general case will not
concern us here.

A very interesting special case is when the system starts in a pure
state (as in Eq \ref{sepu}) and evolves into a mixed state in the
following way:
An initial state
\begin{equation}
|\psi(t_1)\rangle_w = \left( \sum\sqrt{p_i}|i\rangle_s \right) \otimes
|\chi \rangle _r
\label{dc1}
\end{equation}
evolves into
\begin{equation}
|\psi(t_2)\rangle_w =  \sum\sqrt{p_i}|i\rangle_s \otimes |i\rangle_r .
\label{dc2}
\end{equation}
(In Eqs \ref{dc1} and \ref{dc2}   ${ }_s\!\langle i|j\rangle_s = {
}_r\langle i|j\rangle_r = \delta_{ij}$.)
The probabilities assigned to states $|i\rangle_s$  do not change, but
the  time evolution correlates each state  $|i\rangle_s$ with its own
orthogonal state
$|i\rangle_r$ in the environment.
A comparison of Eq \ref{dc2} with Eq \ref{sumsepu} shows that at time
$t_2$ the  $|i\rangle_s$'s
form the Schmidt basis for the system.

As an illustration, one can think of a pendulum, with the surrounding
air as the environment.  If one could imagine setting up the pendulum
in some huge number eigenstate (corresponding to a macroscopic
energy) the pendulum would initially be ``spread out'' over a
macroscopic region.  However,  the interactions with the air would
rapidly correlate the
state of the air molecules with the position of the pendulum.  Such a
process corresponds to the evolution indicated by equations \ref{dc1}
and \ref{dc2}.
The timescale for this process (which is related to the
time between collisions between the pendulum and air molecules)
is
many orders of magnitude  faster than the oscillation time of a
macroscopic pendulum \cite{wz86}.

In order to simplify the following discussion, I will assume the
system itself is static, and consider only the effects of the
interactions with the environment.
Because the ``correlation timescale'' and the system dynamics
timescale are usually so different, this simplification is quite valid
over a range of (intermediate) times.  Furthermore, the discussion
which follows
can be easily generalized to include the system dynamics.

\subsection{Impact of correlations on coherence}

\label{iococ}

The presence of correlations chooses a special basis for the subsystem
(the Schmidt
basis).  Knowledge of the probabilities $p_i^S$ assigned to these
basis states
is all one
needs in order to calculate {\em any} amplitudes (as shown in Eq
\ref{expx}).

The correlations can have different implications, depending on the
circumstances:

\bigskip

{\bf Case I:} If initially one is given the the $p_i^S$'s
(probabilities in the basis which {\em becomes} the Schmidt basis once
correlations are established) , and one is
later going to
measure the system in the Schmidt basis, then one will not be able to
tell whether the correlations are present.  The results will be the
same with or without correlations.  In this case one is always working with
the $p$'s and one thus does not observe quantum coherence.

{\bf Case II:}  If one started with the probabilities $p_i'$'s
assigned to some other (primed) basis, the onset of correlations could
greatly reduce the value of this information.  One would essentially need the
$p_i^S$'s to calculated anything after the correlations were
established (which in turn depend on the probability {\em amplitudes} in
the primed basis, not just the $p_i'$'s).

{\bf Case III:} If one again starts with the $p_i^S$'s
(probabilities in the basis which {\em becomes} the Schmidt basis once
correlations are established) but {\em thinks} that the system is isolated,
then if one later measures in some other basis, one will {\em expect}
to see quantum coherence effects.  One would need to know the initial
{\em amplitudes } to determine.
\begin{eqnarray}
p_x &=& \sum_{i,j} \alpha^{\star}_j \alpha_i { }_s\!\langle x |
i\rangle_s^S { }_s^S\!\langle j | x
\rangle_s \\ & \neq & \sum_i p_i^S \left|{ }_s\langle x| i\rangle_s^S
\right| ^2. \label{correlatedway}
\end{eqnarray}
If the system in actually {\em not} isolated and  gets correlated with the
environment (in the form
given by Eq \ref{dc2}) then Eq \ref{correlatedway} gives the correct
answer, and the anticipated effects of quantum coherence are not
present.  For this reason the setting up of correlations is often
called ``loss of quantum coherence'' or ``decoherence''.

\subsection{Some remarks on decoherence}

\label{srod}

If the world is fundamentally quantum mechanical, why did it take us
so long to discover quantum coherence?  After all, people have studied
physical phenomena over most of human history without any need for
the notion of quantum coherence.
I believe that the primary reason is that there is very good agreement
(among observers and environments) as
to what basis one uses both to measure and describe the world --
namely a basis which closely approximates wave packets fairly
localized in $x$ and
$p$.  Since nice heavy  macroscopic things tend to have WKB evolution,
the only  quantum coherence effects have a natural interpretation in terms of
a classical momentum, as long as one sticks to
this ``classical''
basis (see Section \ref{aiws}).
Furthermore, since the environment
gets correlated with things in the same basis this corresponds to Case
I in Section \ref{iococ} and the effects of decoherence are unimportant.
For these reasons
the effects of quantum coherence are definitely not part of everyday
experience, and it takes the efforts of a
clever experimentalist to set up a situation in which such effects are
important.
(I suppose one could just as well say
we {\em did} discover the complex phase of the wavefunction ages ago, but
we interpreted it as the  momentum!)

There are various interesting things one could say about this
``conspiracy'' between
us and the environment to use the same measurement basis.  On one
hand, the action of the environment gives a tremendous evolutionary
disadvantage to creatures which might choose to measure in some other
basis\cite{z91,paz&habib&zurek93r}.
The information acquired doing unusual measurements  is
rapidly rendered useless buy the decohering effects of the environment
(see Section \ref{iococ} Case II).  On the other hand, given the laws of
physics which govern creatures and environments alike, nature probably
could  not have easily made the two interact so differently.

It is worth mentioning here that there are a great variety of physical
situations which might be regarded as interactions between
a ``system'' and ``environment''.  Only a small number of idealized
cases have
been analyzed carefully (for example
\cite{z82,walls&milburn85,j&z85,u&z89}.
Furthermore, the conditions
under which an environment can cause some decoherence without
destroying the
quantum coherence which is necessary for classical WKB evolution are not
always simple ones\cite{paz&zurek94}.
A really solid understanding of the extent to which everyday physical
processes select a special basis will only emerge after a
more systematic effort.

Another view is that even without the environment, our choice of
interaction basis
and the WKB classicality are enough to account for our classical
perspective
on the world (see Section \ref{iococ} Case I).  Thus decoherence need
not be  mentioned when discussing why the world looks so classical
to us.

In practice, it is usually hard to separate the action of the
environment from the act of measurement.  For example, when we see
something we
are counting on the presence of ambient light, which will act as a
decohering environment regardless of whether we choose to ``bleed
off'' a few photons for the purpose of measurement.

In any case, an environment which actively defines a special basis
through decoherence is present in a wide variety of situations.  It is
essential that all such decohering effects are properly accounted for
when one is looking for quantum coherence effects.  In both examples
mentioned here (cosmic squeezed states and quantum computers) there
are a great many environmental effects which will tend to
destroy the quantum coherence.
(See
\cite{j&z85} for a nice series of illustrations.  There it shown that
even the microwave
background photons have a strong decohering effect on macroscopic
objects.  For an application of the ideas of decoherence to quantum
optics see, for example, \cite{walls&milburn85}.)

\section{Quantum computers}

\label{qc}

\subsection{Preliminaries}

\label{p}

Let us talk about computers in terms of the space of
possible computational states.  From this point of view, the physical
``computer'' consists of every possible bit which is relevant to the
computer's operation, viewed as  simple two state quantum system.
In this idealization, the  states evolve in discrete time steps (from
one computational cycle to the next) according
to the rules by which the computer is designed to run.  (This evolution
will not be unitary    unless the computer is ``reversible'', something
real computers essentially never are.)

To illustrate some points it will be useful to consider the following
simple-minded ``copycat'' computer.
This computer has two bits of memory, and its
rules of operation are simply to copy the contents of register 1 into
register 2 and then remain static for all remaining time steps.  All
possible examples of this computer's evolution
are given by Table \ref{cattable}.
\begin{table}
\begin{tabular}{c|l|l|l|l}
Time step & \multicolumn{4}{c}{Computer state }\\ \hline
0 & $|0\rangle_1 \otimes |0\rangle_2 $  & $|1\rangle_1
\otimes |0\rangle_2 $  & $|1\rangle_1 \otimes |1\rangle_2 $
&  $|0\rangle_1 \otimes |1\rangle_2 $ \\
& & & & \\
1 & $|0\rangle_1 \otimes |0\rangle_2 $  & $|1\rangle_1 \otimes
|1\rangle_2 $  & $|1\rangle_1 \otimes |1\rangle_2 $ &  $|0\rangle_1
\otimes |0\rangle_2 $ \\
& & & & \\
2...$\infty$ & $|0\rangle_1 \otimes |0\rangle_2 $  & $|1\rangle_1
\otimes |1\rangle_2 $  & $|1\rangle_1 \otimes |1\rangle_2 $ &
$|0\rangle_1 \otimes |0\rangle_2 $
\end{tabular}
\caption{The time evolution of a simple ``copycat'' computer viewed
in the ``computational basis''.}
\label{cattable}
\end{table}

The computer's evolution defines a preferred ``computational'' basis
\begin{equation}
\{ |0\rangle_1\otimes |0\rangle_2,\; |0\rangle_1\otimes |1\rangle_2, \;
|1\rangle_1\otimes |0\rangle_2,\; |1\rangle_1\otimes |1\rangle_2\}
\label{compbasis}
\end{equation}
which I have used above.
To illustrate the disadvantages of other bases I use
\begin{equation}
|\pm\rangle \equiv { |0\rangle \pm |1\rangle \over \sqrt{2} }
\label{psdef}
\end{equation}
and the defining rules given by Table \ref{cattable} to
arrive at Table \ref{cattableb}.
\begin{table}
\begin{tabular}{c|l|l}
Time step & \multicolumn{2}{c}{Computer state }\\ \hline
0 & $|+\rangle_1 \otimes |+\rangle_2 $    & $|-\rangle_1 \otimes
|-\rangle_2 $
 \\
 & & \\
1...$\infty$ & ${1\over 2}(|+\rangle _1 |+\rangle _2 \,+\, |-\rangle _1
|-\rangle _2 )$  &  $0 $ \\ \hline \hline
0  & $|-\rangle_1
\otimes |+\rangle_2 $
&  $|+\rangle_1 \otimes |-\rangle_2 $ \\
 & & \\
1...$\infty$  & $
{1 \over 2}(|+\rangle _1|+\rangle _2  \,+\, |-\rangle _1 |-\rangle _2  )$  &
$0 $
\end{tabular}
\caption{The time evolution of the copycat computer, viewed in
another basis (the non unitarity of the evolution is quite apparent).}
\label{cattableb}
\end{table}

Here are two (related) ways in which the evolution of a computer
is simple in the computational basis:

{\em i)}
If the initial state is exactly one of these basis states (not a
superposition),
the expansion in the computational basis of the state at later times
will always contain only one term.   That means that if arbitrary initial
conditions are expressed in the computation basis, and one later
measures in the computational basis, no quantum coherence effects will
be observed.

{\em ii)} If the initial state is exactly a computational basis state
then the states of all individual ``bit'' subsystems remains pure
throughout the evolution. (This follows from {\em i)} and the fact
that the all computational basis states are simple tensor products of
the individual bit states.)  When starting with a non-computational
basis state (See Table 2),
one can say that the other parts of the computer act as an environment
which ``decoheres'' the individual bits.

Having established the importance of the computational basis, one can
now discuss the notion of quantum computation.  Quantum computation
means many things to many people.  The question relevant to the
discussion here is\cite{feynman85,deutsch85,deutsch89}:  Can any
advantage by gained by running a computer
in a coherent superposition of computational basis states?  This
subject was pioneered by Feynman, and recent advances by Deutsch have
greatly enlivened the field\footnote{
Since one can solve the Schr\"{o}dinger equation on a classical
computer there is in principle {\em nothing} one can calculate on a
quantum
computer which can not also be calculated on an ordinary computer.
The growing interest in quantum computation has to do with the fact
that quantum computers can calculate certain things much faster.
}.

\subsection{Remarks on quantum computers}

\label{roqc}

The following remarks illustrate the many links between the field of
quantum computation and the discussion in this paper.  An expert in
quantum computation  will find nothing new here.

{\bf Point  I)} Typically, if one were in  the end to measure a quantum
computer in the
computational base there is no  point in
starting in a coherent superposition.  It is just the
{\em probabilities}  assigned to the initial computational basis
states which will figure into the end result, and there are no quantum
coherence effects.

For example, if one starts the copycat computer in
the state  $2^{-1/2}(|0\rangle_1 \otimes |0\rangle_2 +|1\rangle_1
\otimes |0\rangle_2 ) $, and later measures it in the computational
basis, the results would be indistinguishable from simply using the
a classical
random number generator to choose initial states $|0\rangle_1 \otimes
|0\rangle_2 $ or $|1\rangle_1
\otimes |0\rangle_2 $ with equal probability\footnote{Things might not
be so simple if two coherent components evolved to the same final
state, and the computer varied the phases of the two coherent
components differently  along the way.  (This would not be an issue for
a reversible computer.)
}.
Much of the recent excitement in the field comes from the realization
that if one does measure in some basis other than the computational
basis one can achieve interesting results.

{\bf Point II)}  Having decided to measure the final computer state in some
other basis, one has to be sure to keep the interactions with the
environment from destroying the quantum coherence
(that is, look out for Case {\em III} in section
\ref{iococ}).  Unfortunately, the decohering effects of the environment
are widespread, and hard to avoid.  Fortunately,  (unlike the case of
squeezed states in the early universe) observing quantum coherence in
a computer is in principle an engineering problem which might actually
be solved by a sufficiently creative design  (see for example \cite{lloyd93}).

{\bf Point III)}  The extent to which a quantum computer (or anything else
for that matter) looses quantum coherence is at least formally a clearly
quantifiable thing: If one has an accurately evolved state for
$|\psi\rangle_w$ (in tensor product space ``{\em
computer}$\otimes${\em rest}'') and one hopes to find the computer in
the coherent superposition
\begin{equation}
\alpha_1 |1\rangle_c + \alpha_2 |2\rangle_c,
\label{purecstate}
\end{equation}
then one should simply calculate the off  diagonal density matrix element
\begin{equation}
\langle 2 | \rho_c | 1 \rangle \equiv \langle 2 | \left( \tr _r
|\psi\rangle_w { }_w\!\langle \psi|  \right)  | 1 \rangle.
\end{equation}
To the extent that
\begin{equation}
\langle 2 | \rho_c | 1 \rangle = \alpha_1^{\ast}\alpha_2
\end{equation}
(which corresponds to the pure computer state given by Eq
\ref{purecstate}) the coherence has been preserved.  If correlations
with the environment have completely destroyed the coherence one will
get
\begin{equation}
\langle 2 | \rho_c | 1 \rangle =0.
\end{equation}
It is via these off diagonal matrix elements that information about
the phases of the $\alpha$'s (e.g. $\alpha_1^{\ast}\alpha_2$) can
enter into the final answer.

Of course the trick is to get ahold of an accurately  evolved
$|\psi\rangle_w$.  The environment is very big, and one might easily
forget to include a very weak environment-computer interaction which
could still be strong enough to destroy quantum coherence.

\section{Squeezed states in cosmology}

\label{ssic}

Squeezed states are examples of WKB wavefunctions which are highly
spread out in {some} direction in classical phase space (but not
necessarily in {position}).  Despite the classical properties of
WKB states there is thus the possibility of quantum coherence
being  observed  if a basis other than a classical one
is used\cite{schleich&wheeler87,zhu&caves90}.

Grishchuk, and Grishchuk and
Sidorov\cite{grishchuk91q,grishchuk92,grishchuk&sidorov90}
have pointed out that a period of
cosmic inflation in the early universe will cause matter to  enter a
squeezed state. Can this lead to any quantum coherence effects?  As I
have discussed throughout this article, the answer to this question
depends on in what basis one measures the matter, how quickly
the coherence is lost due to environmental  effects, and whether one wants
to call the WKB relation $p(x)=\hbar \partial_xS(x)$ a quantum coherence
effect.

It appears
that in this case the ``environment'' and the ``measurement''
are closely
related (much as discussed at the end of section \ref{srod}), and
these both measure the classical field variables (such as the
gravitational potential). In any case,  previous claims that an
observable quantum coherence effect
had been calculated have turned out to refer only to effects which have a
natural interpretation as classical physics
\cite{albrecht&ferreira&joyce&prokopec93}.

None the less, it has yet to be argued conclusively that no interesting
quantum coherence effects can
be observed, and it may be  worth considering this question
more thoroughly.

\section{Consistent histories}

\label{ch}

The entire discussion in this article is based on a very standard
quantum formalism.  Namely, the squares of amplitudes give
probabilities at a moment in time.  If one wants to know probabilities
at another time one evolves the amplitudes forward (according to the
Schr\"{o}dinger equation) and squares again.

An alternative formulation of quantum mechanics  (pioneered by
Griffiths\cite{g84} and further developed by Omnes\cite{o88a,o88b,o88c,o92}
and invented independently
by Gell-Mann and
Hartle\cite{g-m&h90,jh90,g-m&h91})
assigns probabilities to {\em histories} rather than states
at a single moment in time (as long as certain conditions are met).

In other publications\cite{a91t,a93q} I have made the point that the
standard formalism seems better adapted for a particular sort of problem.
In contrast, I believe
the subject matter of this paper can be discussed quite
elegantly in  the consistent histories formalism.
I have stuck  to the standard formalism in order to reach a wider
audience, and I will only sketch possible links between the two
formalisms here.

I have discussed how (in the absence of of the wrong kind of
decoherence effects) there always is some basis (at later times) to
which the initial probabilities are assigned.  If one always works in this
basis one would never observe the effects of quantum coherence.
If one follows this
special basis throughout time, then one has a set of ``histories''.
If one were assigning probabilities to histories, it would seem natural
to assign the initial probabilities $p_i$ to these histories.  These
``constant $p_i$'' histories probably  do coincide with consistent
histories, and this suggests that the whole content of this paper could be
re-expressed quite nicely in the consistent histories formalism.

\section{Conclusions}

\label{c}

Quantum coherence occurs when one must use probability amplitudes, rather
than just plain probabilities to describe the evolution of a system.

In this paper I have given a number of illustrations of how issues
of quantum coherence come up in different circumstances.
I have examined the double slit experiment,  WKB type systems,
quantum computers, and matter in squeezed states (a special case of
WKB systems).
For WKB systems,  the crucial role played by the complex phase in
relating the quantum Hilbert
space to the classical phase space allows one to take the unusual view
that all of classical physics is a ``quantum coherence effect''.
Special consideration was given  to the ability of
interactions with the environment to destroy quantum coherence.

I have emphasized how the presence or absence of quantum
coherence has to do with what bases one uses to describe
the initial conditions, and to measure the system.
To the extent that these bases are a matter of choice, the
existence of quantum coherence is very subjective.

Realistically, practical limitations on the possible types
of measurements combined with the decohering effects of the
environment can prevent this subjectivity from being realized.
These limitations are the reason why the notion of quantum coherence
has only recently been required, despite our long history of observing
the physical world.  These limitations also provide major (though not
always insurmountable) barriers to the discovery of new quantum
coherence effects.

\section{Acknowledgments}
This work is based in part on a talk presented at one of an ongoing
series of
workshops  on
Quantum Computation  held in Turin and  sponsored by th ISI Foundation
and ELSAG-Balley.
I would like to acknowledge the stimulating environment and generous
support provided by the sponsors.  I have also benefited a great deal
{ }from discussions with Peter Knight on the numerous ways this material
overlaps with work in the field of quantum optics.

\appendix
\section{The Schmidt Decomposition}
\label{tsd}

\subsection{Proof}
\label{proof}

Here is a brief proof that the Schmidt decomposition may always be
performed:
Consider a state $|\psi\rangle$ in a vector space which we choose to
regard as a tensor product space.  Let $\{|i\rangle_1\} $ and
$\{|j\rangle_2\} $ each be some orthonormal
basis in the corresponding subspace.
There always exist $\alpha_{ij}$'s such that
\begin{equation}
|\psi\rangle = \sum_{ij} \alpha_{ij} |i\rangle_1 |j\rangle_2.
\label{a1}
\end{equation}
Furthermore, one can define
\begin{equation}
|\tilde{i}\rangle_2 \equiv \sum_j \alpha_{ij}|j\rangle_2
\label{itwid}
\end{equation}
So that one can always write
\begin{equation}
|\psi\rangle = \sum_{i} |i\rangle_1 |\tilde{i}\rangle_2.
\label{uns}
\end{equation}
In general, the $|\tilde{i}\rangle_2$'s will not be orthogonal or
normalized.

Now consider the special case were the  $\{|i\rangle_1\}$ are the
(normalized)
eigenstates of $\rho_1$ ($\equiv \tr_2(|\psi\rangle\langle\psi |) =
\sum_{ijk} \alpha^{\ast}_{ij}\alpha_{kj} |i\rangle_1\,{ }_1\!\langle k |$),
call them  $\{|i\rangle_1^S\} $.  In this
case the  $\{|\tilde{i}\rangle_2\}$ {\em must} be orthogonal, because
we must have
\begin{equation}
{ }_1^S\!\langle i | \rho_1 | j \rangle_1^S =
{ }_1^S\!\langle i |
\left(\sum_{k,l}|k\rangle_1^S { }_2\!\langle
\tilde{k}|\tilde{l}\rangle_2
{ }_1^S\!\langle l | \right)| j \rangle_1^S =
 { }_2\!\langle \tilde{i}
|\tilde{j} \rangle_2 \propto \delta_{ij}
\label{ax}
\end{equation}
One can then see that the $|\tilde{i}\rangle_2$'s must be eigenstates
of $\rho_2$ :
\begin{equation}
\rho_2 \equiv \tr_1(|\psi\rangle\langle\psi |) = \sum_i
|\tilde{i}\rangle_2\;{ }_2\!\langle\tilde{i}|.
\label{trr2}
\end{equation}
Finally, one notes that the non-zero eigenvalues of both
$\rho_1$ and $\rho_2$ are both given by $p_i =
{ }_2\!\langle\tilde{i}|\tilde{i}\rangle_2\ $, and one can construct
the normalized states:
\begin{equation}
|i\rangle_2^S \equiv (p_i)^{-1/2}|\tilde{i}\rangle_2.
\label{def2s}
\end{equation}
Equation (\ref{uns}) then becomes
\begin{equation}
\left|\psi\right\rangle = \sum_i \sqrt{p_i}\left|i\right\rangle_1^S
\left|i\right\rangle_2^S,
\label{sres}
\end{equation}
which is the quoted result.

\subsection{Remarks}

Here is a remark which often helps people develop some intuition about
the Schmidt decomposition:
If one is given a particular vector in
a vector space, and is allowed complete freedom to choose a basis, one
can always choose a basis in which the expansion of the particular
vector has but one term.  One simply chooses the first basis vector
proportional to the state in question.  To get a complete basis, one
then constructs an
orthonormal set around that first basis vector (using the
Gram-Schmidt orthogonalization procedure).  If one does not have
{\em complete} freedom to choose a basis, but is allowed to choose any
bases within two pre-determined subspaces, then it should not be
surprising that in general one can not get  down to a
single term in the expansion.  However, one should be able reduce the
number of terms, since there is some remaining flexibility, and that is
what the Schmidt form does.  Note that the number of terms in Eq
(\ref{sres}) is equal to the {\em minimum } of the two subspace
sizes, rather than the {\em product} of the two sizes which would
arise in a typical expansion.

I should also mention the question of degeneracy.  Whenever two
$p_i$'s are degenerate the density matrix eigenstates, and thus the
Schmidt decomposition are not uniquely defined.  All I wish to do here
is assert the view that this fact does not detract from any of the
physical points I make in this paper using the Schmidt decomposition.
The presence of correlations does not specify a unique basis when the
eigenvalues are degenerate,
but otherwise the points I make are still valid.
In \cite{a93q} (Appendix A) I give a detailed discussion of degeneracy
and the Schmidt decomposition.

\end{document}